\documentclass[%
 reprint,
superscriptaddress,
 amsmath,amssymb,
 aps,
showkeys
]{revtex4-2}

\usepackage{graphicx}% Include figure files
\usepackage{dcolumn}% Align table columns on decimal point
\usepackage{bm}% bold math
\usepackage{xcolor}
\usepackage{soul}

\usepackage{multirow}
\usepackage{subfigure}
\usepackage{gensymb}
\begin{document}

\preprint{APS/123-QED}

\title{Unveiling the multi-level structure of midgap states in Sb-doped MoX$_2$ (X = S, Se, Te) monolayers}

\author{Marcos G. Menezes}
 \email{marcosgm@if.ufrj.br}
 \affiliation{Instituto de F\'isica, Universidade Federal do Rio de Janeiro, Caixa Postal 68528, 21941-972 Rio de Janeiro, RJ, Brazil}
 
\author{Saif Ullah}
 \email{ullahs@wfu.edu}
\affiliation{Department of Physics and Center for Functional Materials, Wake Forest University, Winston-Salem, North Carolina 27109, United States}

\date{\today}% It is always \today, today,

\begin{abstract}
In this study, we use first-principles calculations to investigate the electronic and structural properties of MoX$_2$ (X = S, Se, Te) monolayers doped with substitutional Sb atoms, with a central focus on the Sb(Mo) substitution. In MoS$_2$, we observe that this substitution is energetically favored under S rich conditions, where the S$_2$ gaseous phase is likely to be present. This result is compatible with a recent experimental observation in Sb-doped MoS$_2$ nanosheets grown by CVD. A similar behavior is found in MoSe$_2$, but in MoTe$_2$ the Sb(Mo) substitution is less likely to occur due to the possible absence of gaseous Te phases in experimental setups. In all cases, several impurity-induced states are found inside the band gap, with energies that span the entire gap. The Fermi energy is pinned a few tenths of eV above the top of the valence band, suggesting a predominant $p$-type behavior, and gap energies are slightly increased in comparison to the pristine systems. The orbital nature of these states is further investigated with projected and local density of states calculations, which reveal similarities to defect states induced by single Mo vacancies as well as their rehybridization with the $5s$ orbital from Sb.
Additionally, we find that the band gap of the doped systems is increased in comparison with the pristine materials, in contrast with a previous calculation in Sb-doped MoS$_2$ that predicts a gap reduction with a different assignment of valence band and impurity levels. We discuss the similarities, discrepancies, and the limitations of both calculations. We also speculate possible reasons for the experimentally observed redshifts of the A and B excitons in the presence of the Sb dopants in MoS$_2$.
We hope that these results spark future investigations on other aspects of the problem, particularly those concerning the effects of disorder and electron-hole interaction, and continue to reveal the potential of doped TMDCs for applications in optoelectronic devices. 

\end{abstract}

 %Use showkeys class option if keyword display desired

\maketitle

%\tableofcontents

\section{\label{sec:intro}Introduction}

The experimental discovery of graphene and its exceptional properties led to the foundation of a new subject, 2D materials, which is increasing ever since and attracting continuous attention \cite{novoselov2005twoBSCCO, coleman2011two, halim2013rational}. These materials are quite diverse in terms of their physical properties and include metals, semiconductors, band insulators, and even more exotic phases such as superconductors, magnets and topological insulators \cite{xi2016ising, guo2014tuning, chen2009experimental, qu2010quantum, ullah2019exotic, ullah2019non, ullah2019tunable, ullah2019theoreticalB3O3, ullah2019theoretical, ullah2018hydrogenation, cao2018unconventional}. In that landscape, monolayer transition metal dichalcogenides (TMDCs), is a class of materials of particular interest from both fundamental and applied points of view \cite{wallace1947band, novoselov2005twoBSCCO, novoselov2012roadmap, weiss2012graphene, wang2012electronics, geim2013van, chhowalla2013chemistry, xu2014spin}. These materials are composed of a combination of transition metals (M) and chalcogen atoms (X) with chemical formula MX$_2$. When viewed from above, the atoms are organized in a structure similar to the honeycomb structure of graphene, but the M and X atoms sit at different layers. A 2H structure results if the two X layers are identical and a 1T structure is observed if one X layer is rotated with respect to the other, with half of the X atoms sitting on top of hexagon centers. Less symmetrical 1$T'$ structures are also observed, in which the atoms are slightly displaced from their equilibrium positions in 1T. \cite{splendiani2010emerging, mak2010atomically, kong2013synthesis, ruppert2014optical, manzeli-natrevmat-2017, sokolikova-csr-2020}. Within this class, an interesting case is that of semiconducting TMDCs, among which molybdenum disulfide (MoS$_2$) is the most widely studied material. In free-standing form, its most stable phase is 2H and, like many 2D semiconductors, its band gap can be tuned both in magnitude and nature (direct/indirect) by the number of stacked layers \cite{splendiani2010emerging, mak2010atomically}. In particular, the direct gap found in the monolayer makes it much more attractive than the bulk form for applications in optoelectronic devices. In addition to this, these materials display quite fascinating features, such as coupled spin-valley physics and, in the 1$T'$ phase, the quantum spin Hall effect, with potential applications in spintronics \cite{mak2010atomically, mak2014valley, splendiani2010emerging, yu2016valley, zeng2012valley, xiao2012coupled}.

In order to fine-tune the electronic properties of a semiconductor, impurity doping can be considered as an efficient strategy. In fact, the introduction of impurities such as substitutional atoms and defect complexes usually induce localized electronic levels inside the band gap, which can strongly modify the electronic, transport and optical properties of the material. In this regard, a significant effort has been employed to study the doping of various foreign impurities in MoS$_2$. For example, theoretical calculations have addressed the electronic, magnetic and optical properties of defect complexes involving a Mo vacancy and absorbed C, O, Si, Ge and several transition metal atoms \cite{ataca2011functionalization, dolui2013possible, fuhr2004scanning}. Haldar and coworkers have performed a detailed study on the electronic, structural and optical properties of several point defects in TMDCs such as single and double vacancies and interstitial defects \cite{haldar-prb-2015}. Komsa and Krasheninnikov have performed a similar study in monolayer and bulk MoS$_2$ \cite{komsa-prb-2015}, which also considered the effects of growth conditions such as temperature in the energetics. Another study confirmed the origin of a magnetic ground state by replacing a Mo atom with Mn, Fe, or Co impurities \cite{lin2014charge}. Sanvito and coworkers investigated various substitutional dopants on both Mo and S sites and found that the substitution of Nb at a Mo site is suitable for inducing $p$-type conductivity, whereas Re can induce $n$-type behavior \cite{dolui2013possible}. In addition, both $n$, and $p$-type conductivities can be achieved by molecular adsorption \cite{fuhr2004scanning, suh2014doping}. Finally, the doping of Pd and Au atoms has also been reported in MoS$_2$ \cite{choi2014lateral, ma2016adsorption}. 

These theoretical studies were closely followed by experimental reports. For example, a stable $p$-type doping is achieved by Zn and Nb doping in MoS$_2$ \cite{xu2017p,suh2014doping}. Additionally, the optical properties of this material can be fine-tuned with the introduction of Se, W and Er atoms, leading to interesting optoelectronic properties \cite{li2014growth, li2015lateral, dumcenco2013visualization, su2014band, bai20162d}. The synthesis of Co-doped MoS$_2$ has also been realized for field effect transistor applications \cite{li2015synthesis}. In addition, the formation of CoS$_2$ films was also observed within MoS$_2$ grown in high temperature and display half-metallic properties. Doping with Cr, V, and Mn atoms along with various other transition metal atoms in MoS$_2$ also results in interesting electronic and optical properties \cite{gao2016transition, robertson2016atomic, zhang2015manganese}. Sulfur vacancies and Mo antisite defects have been identified in MoS$_2$ through STEM measurements in combination with \textit{ab-initio} calculations. In particular, the calculations reveal that the antisite defects yield a rich multi-level structure inside the band gap, with several defect-induced localized states \cite{hong-natcomm-2015}. More recently, S vacancies and a DX-like center have also been identified in Mo$_x$W$_{1-x}$S$_2$ alloys through a combination of experimental techniques. The DX center was attributed to either a substitutional impurity replacing a cation or a small interstitial atom bonded to neighboring S atoms \cite{tian-natcomm-2020}. Finally, the homogenous doping of Sb in MoS$_2$ has been realized through CVD and a preference for the Sb substitution at the Mo sites was observed \cite{zhong2019electronic}. In this case, small redshifts of the excitonic peaks are observed and were attributed to a reduction of the bandgap. However, despite these very interesting results, the energetics of the defects and the characteristics of impurity-induced levels within the band gap have not been fully addressed for this particular system. 

In this work, we make use of theoretical calculations based on density functional theory (DFT) to study the electronic and structural properties of substitutional Sb doping in 2H-MoX$_2$ monolayers, where X = S, Se, and Te. We aim to carefully describe the formation energies of the defects and the electronic structure of impurity-induced levels inside the band gap. To that end, we have employed calculations with spin-polarization and, in a later step, spin-orbit effects were also included. These materials have been selected in order to illustrate the effects of Sb doping in semiconducting TMDCs, such that similar features should be found, for instance, in 2H-WX$_2$ and other semiconducting structures. The article is organized as follows. In the next section, we present the methods of our DFT calculations. In section \ref{sec:results}, we present and discuss our results by separately considering various important aspects of the problem, such as the electronic properties of the pure systems, structural properties and stability of dopants and the electronic properties of the impurity levels. Our conclusions are presented in section \ref{sec:conclusions}.

\section{\label{sec:method} Methods}

All DFT calculations were performed with the aid of the Quantum Espresso (QE) suite \cite{QE-2009, QE-2017} unless stated otherwise. We use norm-conserving pseudopotentials to model the ion-electron interaction and a PBE-GGA exchange-correlation functional for the electron-electron interaction \cite{perdew1996generalized}. The wavefunctions are expanded in a plane-wave basis with a cutoff energy of 80 Ry. Following a similar recipe to that from Ref. \onlinecite{zhong2019electronic}, we built a $5 \times 5$ supercell of 2H-MoX$_2$ consisting of 75 atoms with a single atom replaced by a Sb atom. For the Sb(Mo) (Sb(S)) substitution, this results in an atomic defect concentration of 4\% (2\%) with respect to the total number of Mo (S) atoms. The Brillouin Zone (BZ) is sampled with a $3 \times 3 \times 1$ Monkhorst-Pack k-point grid \cite{monkhorst-pack}. All atomic positions as well as the cell dimensions were relaxed until the forces on each atom were smaller than $10^{-3}$ Ry/bohr and total energies changed by less than $10^{-4}$ Ry. The target pressure for the variable-cell calculation is set to zero, with a tolerance of $0.5$ kbar. Finally, a vacuum of $20$ \AA \ is included in order to eliminate spurious interactions with periodic images. The effects of spin polarization are included in each case, but all systems relax to a non-magnetic ground state. Additional calculations were also performed with the VASP code, at PBE-GGA level with PAW pseudopotentials and a plane-wave cutoff of 550 eV, in order to provide further support to our main results \cite{kresse1994ab,perdew1996generalized,blochl1994projector,kresse1996efficiency}.We also consider spin-orbit coupling (SOC) calculations with both QE and VASP codes. The results of most of these calculations are shown in the supplemental file and discussed throughout the text.

\section {\label{sec:results} Results}

\subsection{\label{subsec:pure} Electronic and structural properties of pristine 2H-MX$_2$}

\begin{figure*}[ht]
    \centering
    \includegraphics[width=\textwidth]{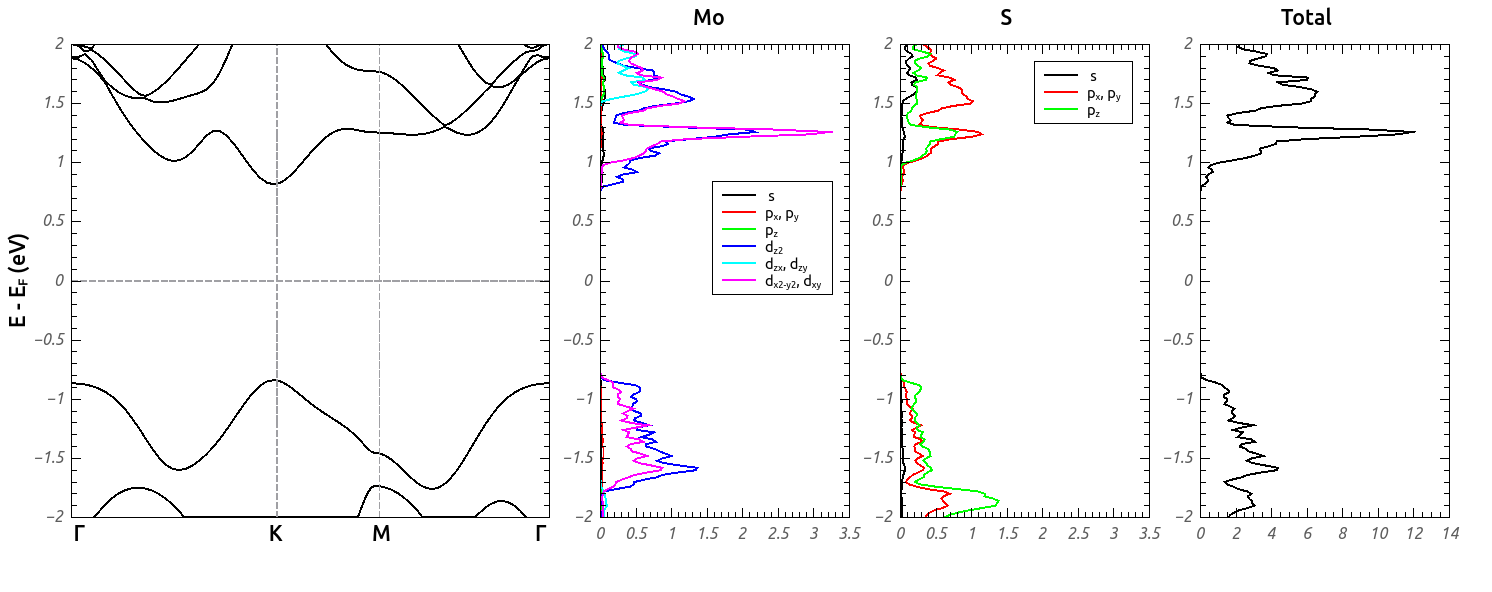}
    \includegraphics[width=\textwidth]{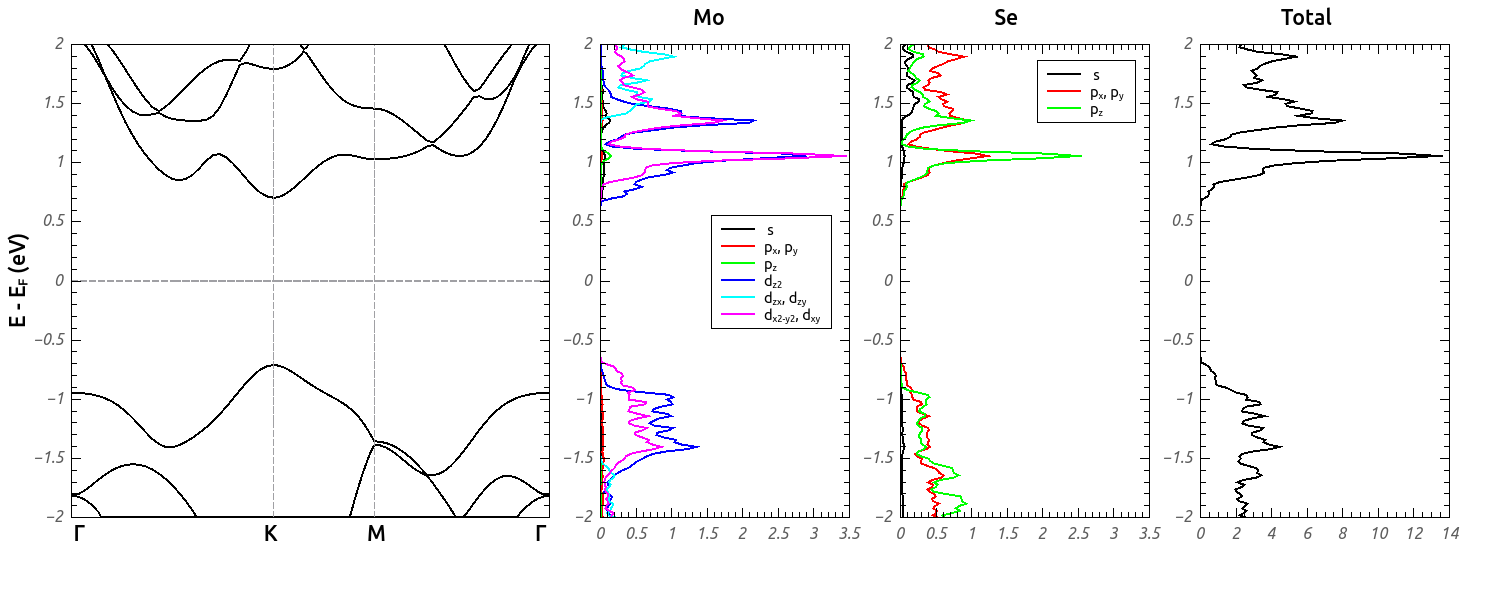}
    \includegraphics[width=\textwidth]{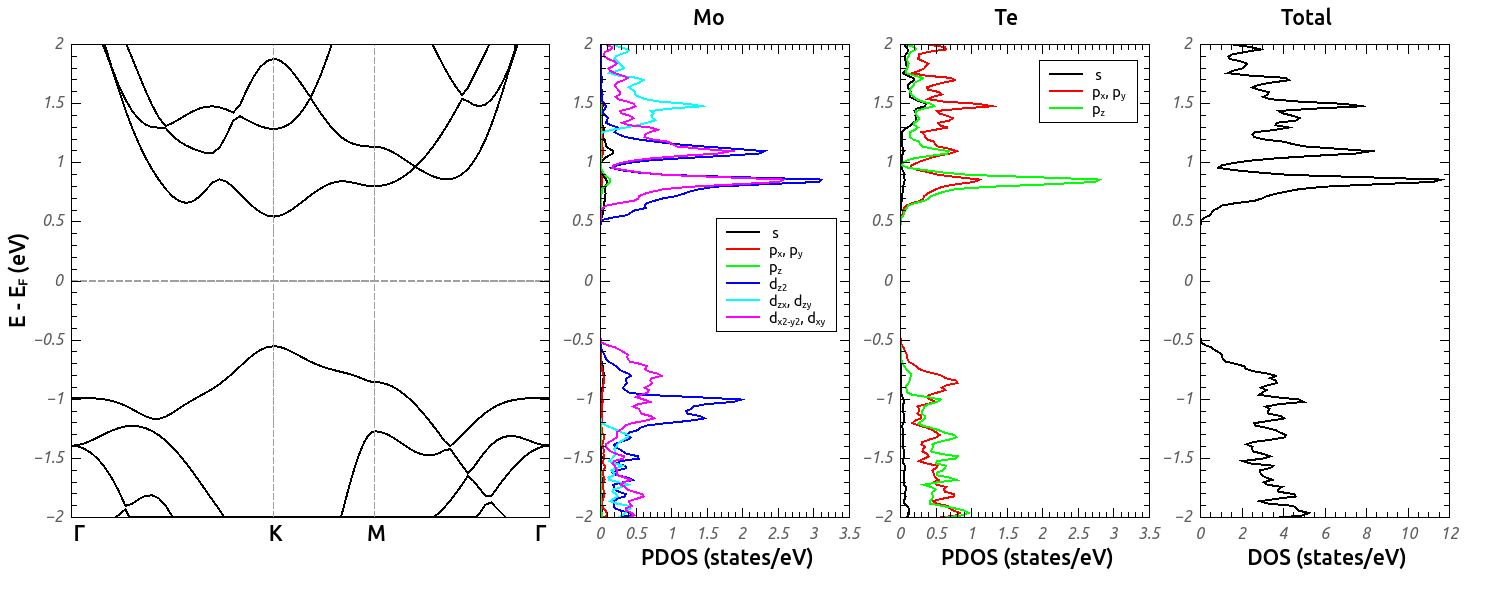}
    \caption{\label{fig:bands_pure} Band structure, projected density of states (PDOS) and total DOS of pristine MoX$_2$ in the absence of spin-orbit coupling. All energies are measured with respect to the Fermi energy. The contributions from each atomic orbital are colored according to the legend of each column. For the X atoms, the plots show the sum of the contributions from the two atoms in the unit cell.}
\end{figure*}

We begin by recalling some properties of pristine 2H-MoX$_2$, as they will be central to our discussion of the electronic properties of the doped structures. In Table \ref{tab:structure_pure}, we present the equilibrium properties of the three materials as obtained from our calculations. The calculated lattice constants, bond lengths and thicknesses of the monolayers are in excellent agreement with previous reports \cite{ataca2011functionalization, yandong-pccp-2011, zollner-prb-2019, wang-jpd-2013, haldar-prb-2015}. In Fig. \ref{fig:bands_pure}, we present the band structures in the absence of spin-orbit coupling (SOC), together with the projected and total density of states (PDOS and DOS). As we can see, all gaps are direct, with the valence band maximum (VBM) and conduction band minimum (CBM) lying at the K point. Their magnitudes are reported in Table \ref{tab:structure_pure} and agree with previous reports \cite{haldar-prb-2015}. The PDOS calculations show that the low-energy states (close to the Fermi energy) are composed mostly of $d$-orbitals from the Mo atoms and $p$-orbitals from the X atoms. Additionally, the contributions from $d_{xy}$ and $d_{x^2-y^2}$ orbitals of Mo are identical, as are those from the orbitals $d_{yz}$ and $d_{xz}$ of Mo and $p_x$ and $p_y$ of X. This is a consequence of the crystal field of the material, as the 2H-MoX$_2$ structure has the point group symmetry $D_{3h}$ and this group has only one and two-dimensional irreducible representations. The orbitals with identical contributions correspond to basis functions of the same two-dimensional representation and, as such, transform as linear combinations of the same functions when subject to all symmetry operations of the group. Therefore, as the PDOS is calculated through a sum over all k-points in the BZ, all symmetry-equivalent points in a band are included and the sum of the weights of each basis function of the representation over all equivalent points is identical. Moreover, in the case of the X atoms, the orbitals from the top and bottom layers are arranged in symmetrical and anti-symmetrical combinations which are either even or odd with respect to the mirror plane symmetry in the central Mo layer. This symmetry is present at every k-point of the BZ, despite the general symmetry being reduced at arbitrary points.
\begin{table*}[ht]
    \centering
    \caption{\label{tab:structure_pure} Equilibrium properties of pristine MoX$_2$: lattice constant ($a$), Mo-X bond length ($d_{\rm{Mo-X}}$) and thickness of the monolayers ($L$). The band gaps in the absence ($E_g$) and presence of spin-orbit coupling ($E_g^{SO}$), and the energy splitting at the top of the valence band ($\Delta_v^{SO}$) are also included.}
    \begin{tabular}{p{24mm} | >{\centering}p{24mm} | >{\centering}p{24mm} | >{\centering}p{24mm} | >{\centering}p{24mm} | >{\centering}p{24mm} | >{\centering\arraybackslash}p{24mm}}
         \hline
                  & $a$ (\AA) & $d_{\rm{Mo-X}}$ (\AA) & $L$ (\AA) & $E_g$ (\rm{eV})  & $E_g^{SO}$ (\rm{eV}) & $\Delta_v^{SO}$ (\rm{eV}) \\
         \hline
         MoS$_2$  & $3.23$    & $2.45$    & $3.19$    & $1.66$    & $1.58$    & $0.15$   \\     
         MoSe$_2$ & $3.36$    & $2.58$    & $3.38$    & $1.42$    & $1.31$    & $0.19$   \\
         MoTe$_2$ & $3.57$    & $2.75$    & $3.64$    & $1.10$    & $0.96$    & $0.22$   \\
         \hline
    \end{tabular}
\end{table*}

\begin{figure*}[ht]
    \centering
    \includegraphics[width=\textwidth]{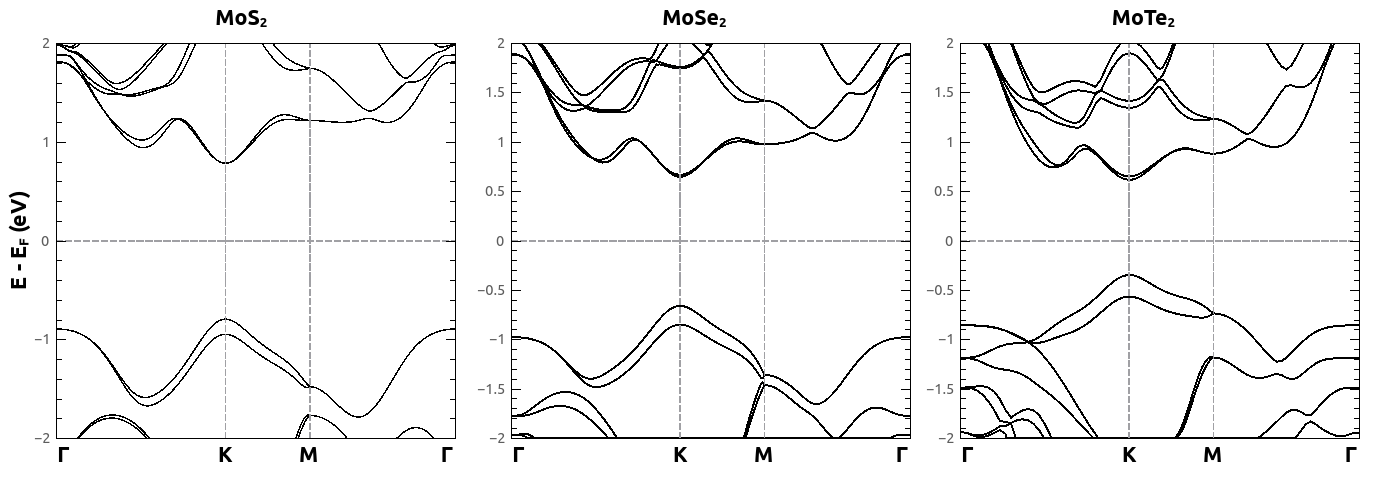}
    \caption{\label{fig:bands_pure_soc} Band structures of pristine MoX$_2$ in the presence of spin-orbit coupling. All energies are measured with respect to the Fermi energy.}
\end{figure*}

The inclusion of SO coupling introduces important changes to the electronic structure, as can be appreciated in Fig. \ref{fig:bands_pure_soc}. The most important effects are a reduction of the band gap and the splitting of the valence band, particularly at the K point. These values are also reported in Table \ref{tab:structure_pure} and, again, an excellent agreement is found with previous reports at similar levels of calculation \cite{kadantsev-ssc-2012, PhysRevB.88.045416, PhysRevX.4.011034, PhysRevB.91.235202, doi:10.1021/acs.jpclett.6b00693} . In particular, the expectation value of the $z$-component of the spin in each of the split bands is opposite at the $K$ and $K'$ points of the BZ, which is a signature of the spin-valley coupling. A similar effect is also observed at the conduction band, but the splittings are much smaller. Our calculated values range from $3$ meV in MoS$_2$ to $35$ meV in MoTe$_2$.

%%%%%

\subsection{\label{subsec:stability} Energetics and stability of Sb-doped structures}

In order to assess the stability of Sb doping, we have evaluated the formation energies of Sb(Mo) and Sb(S) substitutional impurities in MoX$_2$. They are defined as:
\begin{equation}
    \label{eq:formation_eng}
    E_f = E_{def} - E_{pure} - \sum_i n_i \mu_i,
\end{equation}

\noindent where $E_{def}$ is the total energy of a supercell containing a single defect, $E_{pure}$ is the total energy of a pristine MoX$_2$ supercell of the same size and $\mu_i$ and $n_i$ are the chemical potential and number of atoms added ($n_i > 0$) or removed ($n_i < 0$) of species $i$. For the evaluation of the chemical potentials of Mo and Sb, we use as reference structures the bulk bcc phase of Mo and the bulk rhombohedral phase of Sb. For S and Se, we use the isolated diatomic molecule as reference, but we also consider their bulk phases for comparison, as we discuss below. For Te, only the bulk phase is considered. Additionally, the chemical potentials of Mo and X in MoX$_2$ are connected via $\mu_{Mo} + 2 \mu_X = \mu_{MoX_2}$, where $\mu_{MoX_2}$ is the total energy of a pristine unit cell. This allows us to simulate Mo rich (X rich) conditions by setting $\mu_{Mo}$ ($\mu_X$) to the value of the total energy per atom in the reference structure and $\mu_X$ ($\mu_{Mo}$) is calculated via the connecting formula, following a similar recipe to that of Refs. \onlinecite{haldar-prb-2015} and \onlinecite{komsa-prb-2015}.

We begin by discussing the energetics of defects in MoS$_2$, as it is the most widely studied material. From our fully relaxed structures, the formation energies of the Sb(Mo) and Sb(S) defects under Mo rich conditions are $4.68$ and $0.80$ eV, respectively, which suggests that the Sb(S) is the most favorable substitution in that scenario. In contrast, under S rich conditions, the formation energies are $0.72$ and $2.69$ eV, thus reversing the trend. The latter scenario is compatible with a recent experimental observation in which the Sb(Mo) substitution is favored in homogeneously Sb-doped MoS$_2$ nanosheets grown by CVD on a SiO$_2$/Si substrate \cite{zhong2019electronic}. Other growth conditions could also influence this result, such as temperature and the substrate itself. In particular, the formation of single vacancies could also aid the formation of the Sb(Mo) defect. For reference, we have calculated the formation energies of single vacancies in MoS$_2$ at the same level of theory. Under Mo rich conditions, the values we have obtained are $6.95$ eV for a Mo vacancy and $1.27$ eV for a S vacancy.  Under S rich conditions, the formation energies become quite close, with values of $2.99$ and $3.26$ eV, respectively. Therefore, both vacancies are likely to form in this case and could be stabilized by being filled with Sb atoms, with a preference for Mo vacancy sites.

It should be noted that selecting a suitable reference structure for the evaluation of the chemical potential of S under S rich conditions could be a complicated task due to the existence of various S allotropes depending on temperature. For instance, the $\alpha$-S allotrope, which is a solid made of S$_8$ rings loosely bound, can be found below 100\degree C. The liquid-phase can be found beyond 120\degree C and the gas-phase is found by crossing 720\degree C, comprising of numerous S molecules. Nevertheless, S exists as a diatomic gas (S$_2$) above 880\degree C \cite{219851}. Therefore, considering that the CVD method is usually employed for growing quality materials in which the temperature range varies between 650 and 970\degree C, S can be found in gaseous form at typical CVD temperatures \cite{kong2013synthesis,zhan2012large,lee2012synthesis,najmaei2013vapour,van2013grains,komsa-prb-2015}. In that scenario, our results for the energetics of vacancies do not fully agree with those from Ref. \onlinecite{haldar-prb-2015}, in which the S vacancy is favored in both Mo and S rich conditions. This difference can be attributed to a different choice of reference structure for S in that calculation, namely, the bulk crystal. By using as reference the bulk structure of S instead of the S$_2$ gas phase in our calculations, we find formation energies of $2.39$, $1.35$, $4.66$ and $2.42$ eV for Sb(Mo), Sb(S), Mo vacancy and S vacancy, respectively, under S rich conditions. The values for the vacancies agree with those from Ref. \onlinecite{haldar-prb-2015}, thus highlighting the importance of the choice.  Similar outcomes for vacancies in MoS$_2$ are reported in Ref. \onlinecite{komsa-prb-2015}, which considered both reference structures as the chemical potential of S is varied. However, for experimental growth conditions such as those of Ref. \onlinecite{zhong2019electronic}, where temperatures as high as $955$ K are achieved, the gas phase of S$_2$ should be predominant. This could explain the preference for the Sb(Mo) substitution in the experiment, as supported by our calculations.

Naturally, similar trends should also be observed in defective MoSe$_2$ and MoTe$_2$ layers with appropriate choices of reference structures for the Se and Te atoms. For MoSe$_2$, the formation energy of the Sb(Mo) substitution is $3.98$ eV under Mo rich conditions. By using the gas phase of Se$_2$ as reference, the formation energy of this defect reduces to $0.47$ eV under Se rich conditions, indicating that it is likely to be formed. If we use the most stable bulk phase of Se instead, the formation energy is $2.01$ eV. For MoTe$_2$, the formation of the Sb(Mo) defect is $3.15$ eV under Mo rich conditions. Now, considering the boiling temperature of Te is higher than the temperatures commonly found in (CVD) growth conditions, a gas phase should be absent and we only use its most stable bulk phase as a reference. In that case, the formation energy under Te rich conditions is $2.27$ eV, thus making the Sb(Mo) defect less likely to occur in MoTe$_2$. Finally, we have also cross-checked our calculations with the aid of the VASP code and a good overall agreement is observed for the formation energies. A complete table of the values can be found in the supplemental file (see Table S1). Cohesive energies are also included for comparison. 

\subsection{\label{subsec:stability} Electronic and structural properties of Sb-doped structures}

We now focus our discussion on the Sb(Mo) substitution in MoX$_2$ structures, which was observed experimentally for X = S \cite{zhong2019electronic}. We begin with the structural properties of the doped systems. Our variable-cell relaxations indicate that the introduction of the Sb induces important structural distortions in its environment. In Table \ref{tab:structure_doped}, we report a few parameters that describe these modifications. First, notice that the average lattice constant, defined as the lattice constant of the supercell divided by its dimension (5), is slightly increased in all cases in comparison with the lattice constants of the pristine materials. This highlights the importance of a full relaxation, including the supercell vectors. Next, notice that the neighboring atoms move away from the impurity, resulting in an increased Sb-X bond length in comparison with the value for the Mo-X bond in the pristine system. The distance between the first-neighboring X atoms is also increased in comparison with the original value, which is the lattice constant of the pristine system. In contrast, the local thickness of the monolayer, which is the distance between first-neighboring X atoms in opposing layers, is slightly reduced. This thickness increases with the distance from the impurity up to the value found in the pristine system.
In addition, we observe that the trigonal prismatic coordination is preserved in the vicinity of the impurity. This is expected since the impurity does not remove the mirror symmetry found in the original structure, as we discuss in more detail below. Therefore, a structural transition from the 2H phase to the 1T or 1$T'$ phases upon Sb doping is not expected. However, the picture could be different in less symmetrical scenarios, such as different impurity configurations in multilayers and heterostructures. As an example, a recent report indicates that Cu/Co intercalation in multilayer SnS$_2$ induces a stacking-type modification in the structure \cite{gong2018-naturenano}. A similar outcome may result from Sb intercalation in MoX$_2$ multilayers.

\begin{table}[ht]
    \centering
    \caption{\label{tab:structure_doped} Structural properties of Sb(Mo)-doped MoX$_2$: average lattice constant ($a$), defined by the supercell lattice constant divided by its dimension, Sb-X bond length ($d_{\rm{Sb-X}}$), distance between two X atoms which are bound to Sb and belong to the same layer ($d_{\rm{X-X}}$) or opposing layers ($L$). Notice that $L$ corresponds to the local thickness of the monolayer. }
    
    \begin{tabular}{p{16mm} | >{\centering}p{16mm} | >{\centering}p{16mm} | >{\centering}p{16mm} | >{\centering\arraybackslash}p{16mm}}
         \hline
                  & $a$ (\AA) & $d_{\rm{Sb-X}}$ (\AA) & $d_{\rm{X-X}}$ (\AA) & $L$ (\AA) \\
         \hline
         MoS$_2$  & $3.25$    & $2.57$    & $3.53$    & $3.14$   \\     
         MoSe$_2$ & $3.40$    & $2.73$    & $3.74$    & $3.35$   \\
         MoTe$_2$ & $3.60$    & $2.94$    & $4.04$    & $3.58$   \\
         \hline
    \end{tabular}
\end{table}

\begin{figure*}[ht]
    \centering
    \includegraphics[width=\textwidth]{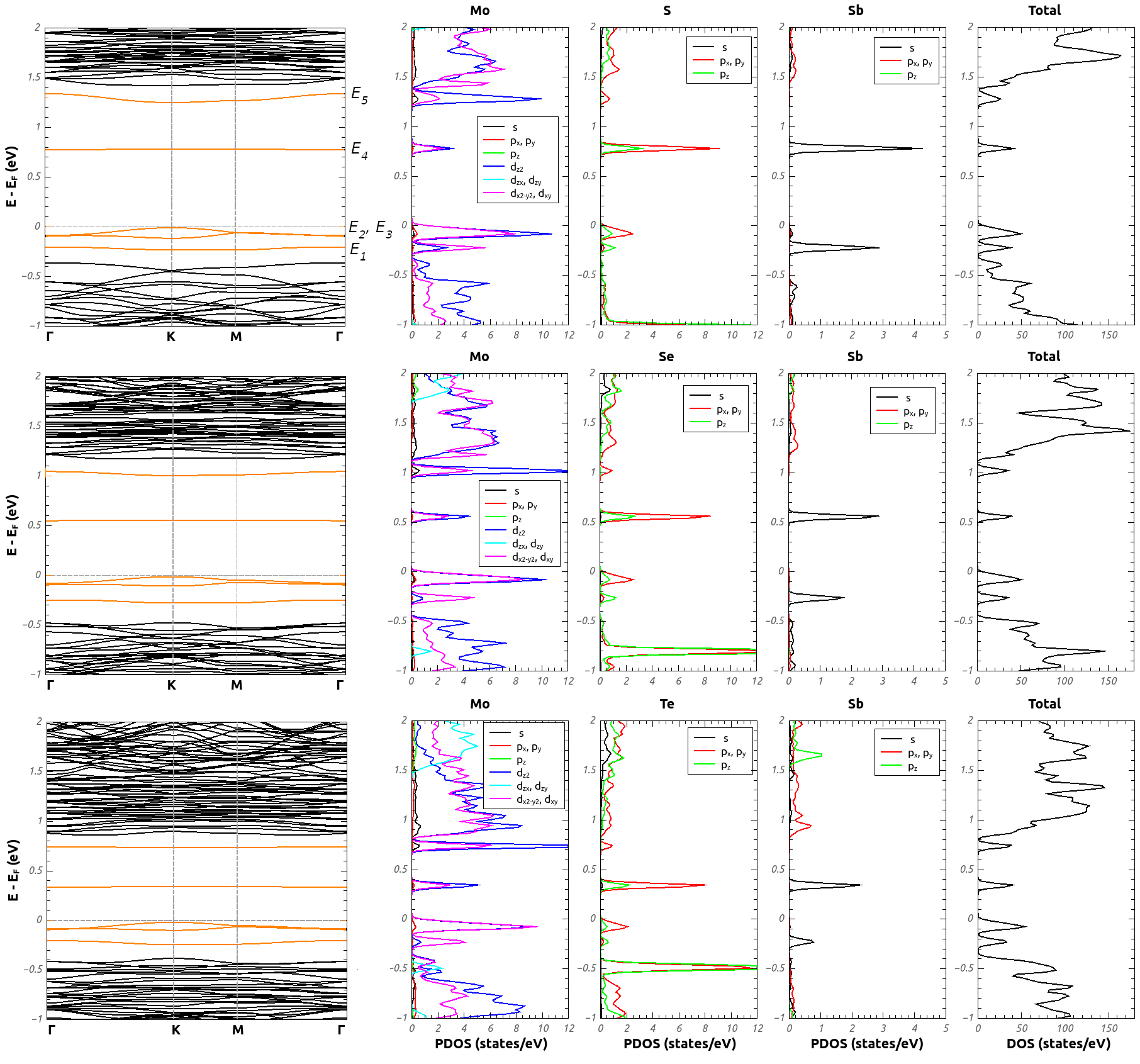}
    \caption{\label{fig:bands_sb} Band structure, projected density of states (PDOS) and total DOS of Sb-doped MoX$_2$ (Sb(Mo) substitution) in the absence of spin-orbit coupling. All energies are measured with respect to the Fermi energy. The impurity levels inside the band gap are highlighted in orange in the band plots and are identified through peaks in the DOS. In the PDOS plots, the X and Mo panels show the sum over contributions from all first and second neighbors of the impurity, respectively.}
\end{figure*}

The resulting band-structures in the absence of spin-orbit coupling are shown in Fig. \ref{fig:bands_sb}, together with the total DOS and the PDOS from the orbitals of the impurity atom and the sum of contributions from surrounding atoms. In the bands, we can clearly see five impurity levels inside the band gap (highlighted in orange), two of which are near-degenerate. Naturally, their signature is also present in the total DOS, where we see four peaks inside the gap and one of them is associated with the near-degenerate levels. We number these levels in order of increasing energy ($E_1$ to $E_5$), measured with respect to the valence band edge, and report the position of the corresponding DOS peaks in Table \ref{tab:energies_sb}. The band gaps and the Fermi energy are also included \footnotetext{Here we report the Fermi energy as given by the calculations, which include the effects of smearing. The actual values, as predicted by electron counting, may differ by about $0.10$ eV, which is width of the smearing.}. Note that the band gaps are slightly increased in comparison with the pristine values from Table \ref{tab:structure_pure}. This result apparently contrasts with the calculations from Ref. \onlinecite{zhong2019electronic}, where a gap reduction is observed and attributed to a possible explanation of the redshift of the excitonic peaks induced by the impurity. However, we believe that both results can be reconciled with a proper assignment of the impurity levels, as we discuss below. Finally, our calculations yield Fermi energies about $0.32$ - $0.42$ eV above the valence band edge, suggesting a predominant $p$-type behavior in a similar fashion to other dopants such as Nb \cite{dolui2013possible, xu2017p,suh2014doping}.

\begin{table}[ht]
    \centering
    \caption{\label{tab:energies_sb} Energies of the impurity induced levels ($E_1 - E_5$), band gap ($E_g$) and Fermi energy ($E_F$) in Sb(Mo)-doped MoX$_2$. All energies are measured from the valence band edge and are in eV.}
    
    \begin{tabular}{p{20mm} | >{\centering}p{20mm} | >{\centering}p{20mm} | >{\centering\arraybackslash}p{20mm}}
        \hline
                     &  MoS$_2$ & MoSe$_2$ & MoTe$_2$ \\
        \hline
        $E_1$        & 0.10 & 0.16 & 0.14 \\
        $E_2$, $E_3$ & 0.24 & 0.34 & 0.30 \\
        $E_4$        & 1.10 & 0.98 & 0.72 \\
        $E_5$        & 1.60 & 1.44 & 1.12 \\
        $E_g$        & 1.78 & 1.65 & 1.24 \\ % From eigenvalues
        $E_F$        & 0.32 & 0.42 & 0.38 \\
        \hline
    \end{tabular}
\end{table}

\begin{figure*}[ht]
    \centering
    \includegraphics[width=\textwidth]{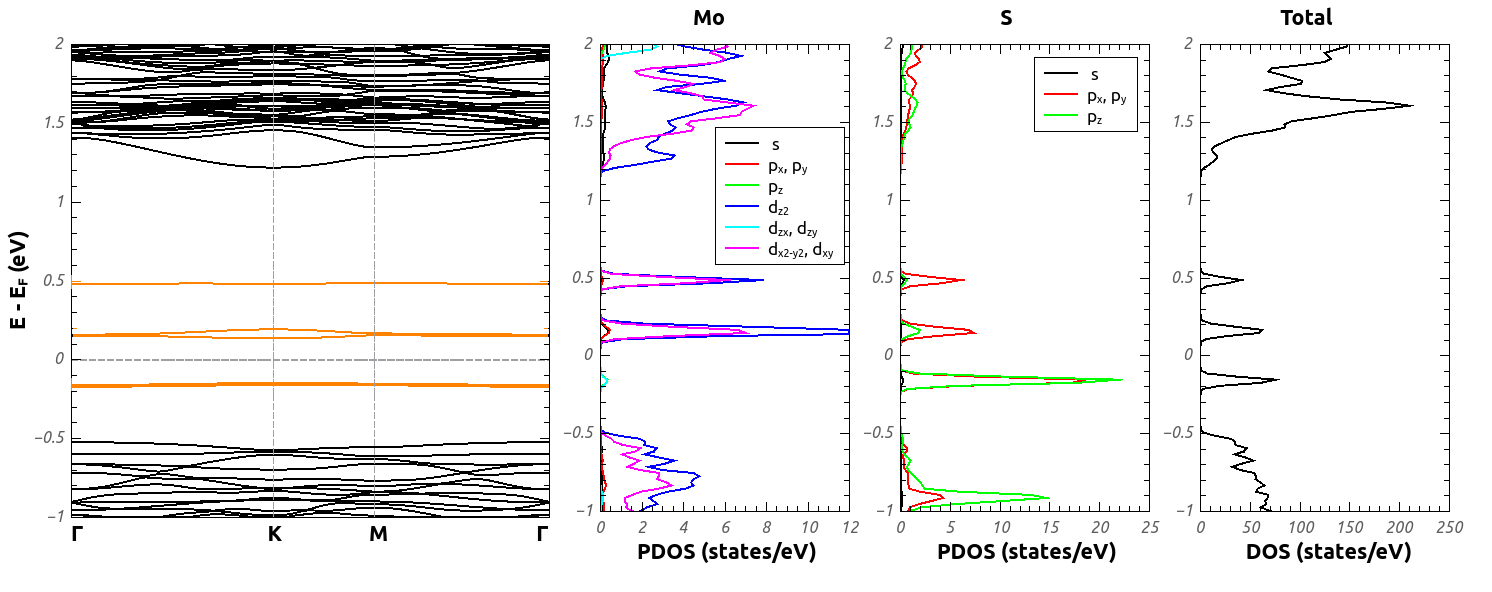}
    \caption{\label{fig:bands_vac} Band structure, projected density of states (PDOS) and total DOS for MoS$_2$ with a Mo vacancy, in the absence of spin-orbit coupling. All energies are measured with respect to the Fermi energy. The colors codes are the same as in Fig. \ref{fig:bands_sb}.}
\end{figure*}

The electronic structure of the impurity levels can further be investigated through the analysis of their wavefunctions. We carry this analysis through PDOS calculations, shown in Fig. \ref{fig:bands_sb}, and a calculation of the local density of states (LDOS) integrated over small energy ranges that span each level or set of levels in the case of near-degeneracy. The LDOS plots for Sb-doped MoS$_2$ are shown in Fig. \ref{fig:ldos_sb}, while those of MoSe$_2$ and MoTe$_2$ are shown in the supplemental file and display similar features (see Figs. S1 and S2). From these plots, we can see that all levels are localized around the impurity and display quite unique features. First, note that only the $E_1$ and $E_4$ levels contain contributions from the $5s$ orbital of Sb itself. On the other hand, all levels contain contributions from $p$ orbitals of neighboring X atoms, but these are particularly stronger in the $E_4$ level. Finally, all levels also contain contributions of comparable magnitudes from $4d$ orbitals of Mo atoms that are second neighbors to the impurity and smaller contributions from more distant neighbors. Interestingly, the $5p$ orbitals from Sb are not involved in any impurity level and, although we have not included the $4d$ electrons of Sb as semicore electrons in our calculations, earlier reports indicate that their contribution is very small in the energy range of interest \cite{zhong2019electronic}. In that sense, the Sb(Mo) substitution effectively reduces the number of available valence orbitals involved in the hybridization, suggesting that some of these levels could display similar features to that of impurity levels induced by a single Mo vacancy. To investigate this possibility, we have calculated the electronic properties of the Mo vacancy in MoS$_2$ at the same level of theory as our Sb(Mo) calculations. The band structure and PDOS/DOS plots are shown in Fig. \ref{fig:bands_vac} and the LDOS plots for the impurity levels are shown in Fig. \ref{fig:ldos_vac}. As in that case, we find five impurity levels inside the band gap, but now we find an additional near-degenerate pair right above the valence band edge. The positions of the corresponding DOS peaks are $0.32$, $0.64$ and $0.96$ eV, as measured from the valence band edge, and agree quite well with previous calculations \cite{ataca2011functionalization, haldar-prb-2015}. The band gap is $1.70$ eV, which is also slightly larger than the pristine value from Table \ref{tab:structure_pure}. If we think of this system as a starting point to understand the electronic properties of Sb(Mo) doping, we may argue that the defect states of the vacancy system rehybridize with the $5s$ orbital of Sb and give rise to the defect states observed in Sb(Mo). By comparing the PDOS and LDOS profiles, it appears that the structure of the second near-degenerate pair in the vacancy system (levels $E_3$ and $E_4$) is preserved by the introduction of Sb, resulting in the near-degenerate pair observed in Sb(Mo) (levels $E_2$ and $E_3$). However, the contribution from the S $p$-orbitals is reduced. The first degenerate pair (levels $E_1$ and $E_2$) and the non-degenerate level in the vacancy system ($E_5$) appear to strongly interact with the $5s$ orbital of Sb, resulting in the remaining levels found in Sb(Mo). In particular, the near-degeneracy of the pair is lifted, resulting in only one level remaining inside the gap, while the other is mixed within the occupied extended states, as we can identify through the PDOS peak at about $-1.0$ eV in Fig. \ref{fig:bands_sb} (top). Notice that this peak moves towards the valence band edge as we change X from S to Te. Simultaneously, the contributions of Sb $5s$ orbitals to levels 2, 3 and 4 are reduced, which suggests that the strength of the interaction diminishes. This could be related to the reduction of the band gap and to an increase of the structural deformation induced by the impurity, as we can see from the increase of the $d_{\rm{X-X}}/a$ ratio in Table \ref{tab:structure_doped}.

\begin{figure*}[ht]
    \centering
    \includegraphics[width=0.9\textwidth]{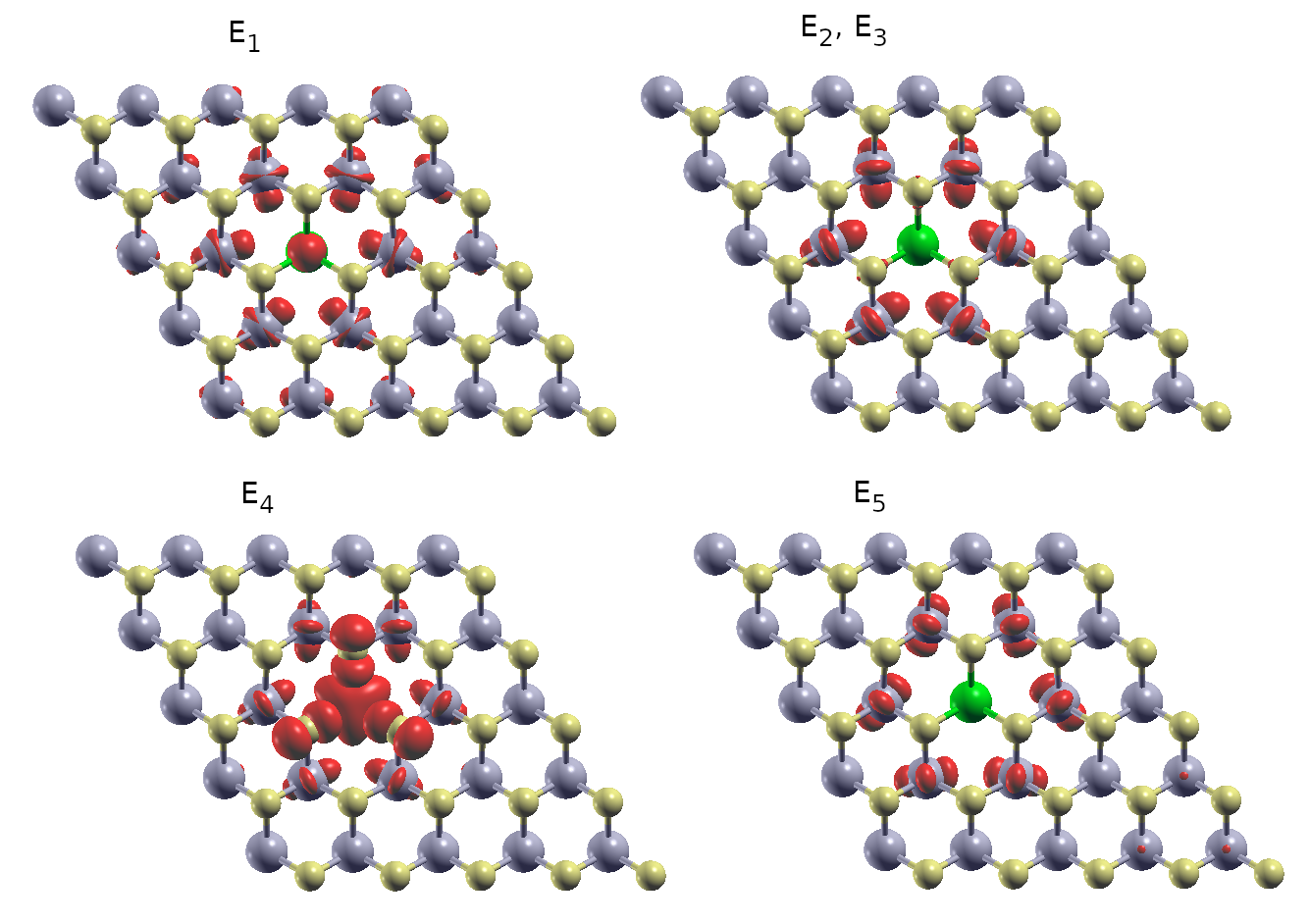}
    \caption{\label{fig:ldos_sb} Local density of states (LDOS) isosurfaces for the impurity levels in Sb(Mo)-doped MoS$_2$. The levels are arranged in increasing energy, as in Fig. \ref{fig:bands_sb} and Table \ref{tab:energies_sb}. The isosurfaces, shown in red, correspond to 10\% of the maximum value found in each case. The Sb impurity is represented as a green sphere, while the Mo and S atoms are represented by gray and yellow spheres, respectively. Similar features are found in the MoSe$_2$ and MoTe$_2$ structures.}
\end{figure*}

\begin{figure*}[ht]
    \centering
    \includegraphics[width=\textwidth]{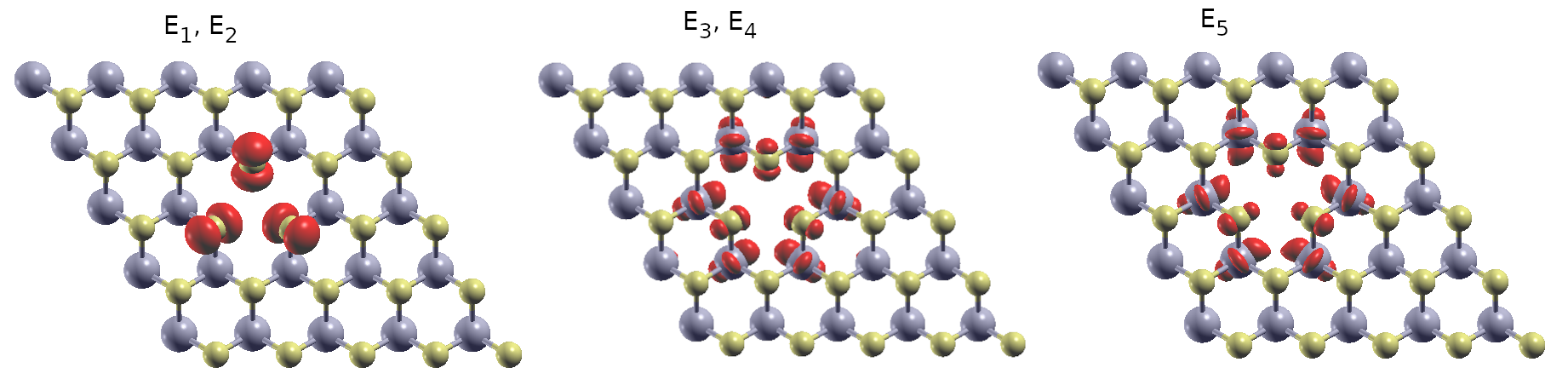}
    \caption{\label{fig:ldos_vac} Local density of states (LDOS) isosurfaces for the impurity levels for a single Mo vacancy in MoS$_2$. The levels are arranged in increasing energy, as in Fig. \ref{fig:bands_vac}. The isosurface parameters and color codes are the same as in Fig. \ref{fig:ldos_sb}.}
\end{figure*}

Another interesting feature of the PDOS calculations in both Sb(Mo) and Mo vacancy systems is that the paired orbital contributions observed in the pristine calculations are preserved. This is a consequence of the fact that the $D_{3h}$ point group symmetry is preserved with respect to the defect site. With the supercell construction, the periodic structure also preserves that symmetry. This property is even observed in the case of Sb(S) doping and the S vacancy, where the mirror symmetry and related operations are lost. In these cases, the symmetry with respect to the defect site is reduced to $C_{3v}$, but the irreducible representations of this group have a similar paired structure of basis functions. The electronic structure of these two cases is briefly discussed in the supplemental file (see Figs. S3 and S4). In particular, the results for the S vacancy agree quite well with a previous report from Ref. \onlinecite{haldar-prb-2015}. Meanwhile, the Sb(S) substitution behaves as an acceptor with a near-degenerate ground state. Naturally, these symmetries will further be reduced in less idealized doping scenarios with disordered configurations, which may lead to an even richer electronic structure. Since such scenarios cannot be address through \textit{ab-initio} calculations, future investigations with semi-empirical models such as the tight-binding method can contribute to increase our understanding of the electronic properties of these defects in MoX$_2$ and other semiconducting TDMCs.

Another important point to address is the effect of the supercell size in our calculations, which allows us to identify possible residual interactions between the impurity and its periodic images. To that end, we have performed a calculation in a larger $10 \times 10$ supercell for two cases of interest, namely, the Sb(Mo) impurity in MoS$_2$ and the Mo vacancy in MoTe$_2$. These situations corresponds to atomic defect concentrations of 1\% with respect to the total number of Mo atoms. A comparison of the band structures for the $5 \times 5$ and $10 \times 10$ supercells can be found in Figs. S5 and S6 of the supplemental material. We also show LDOS profiles for some impurity levels of the larger cell in Fig. S5.
We find that the structure of the impurity levels is preserved, including their energies with respect to the valence band edge and the corresponding wavefunctions. In particular, for the Sb(Mo) defect, the weak dispersion of the $E_1$, $E_2$ and $E_3$ levels observed in Fig. \ref{fig:bands_sb} is eliminated in the larger cell, while that of the $E_4$ is reduced. A similar outcome is observed for the Mo vacancy. These results provide further evidence for the identification of these levels as localized states induced by the impurity. Additionally, they confirm that the main features of these levels, including their energetics and wavefunctions are well described by the smaller $5 \times 5$ cell. We have also performed cross-check calculations with the VASP code for all of the structures considered in this work in the same $5 \times 5$ cell (see Figs. S7 - S10 of the supplemental file) and we find an excellent agreement with the band structures from QE, thus confirming the robustness of our results.

\begin{figure*}[ht]
    \centering
    \includegraphics[width=\textwidth]{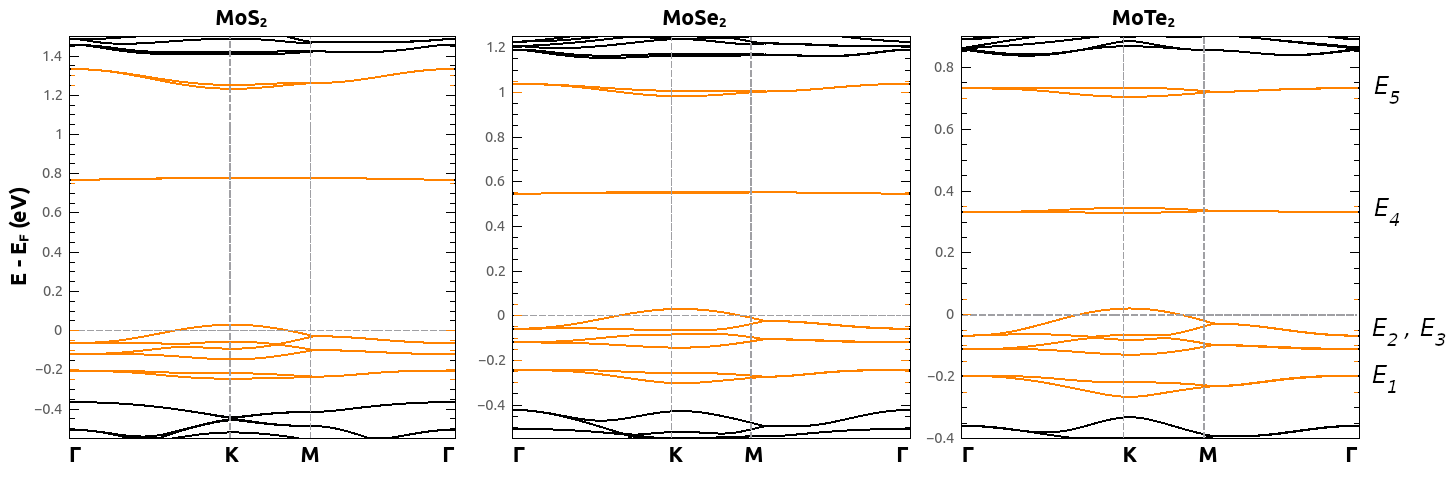}
    \caption{\label{fig:bands_sb_soc} Band structures of Sb(Mo)-doped MoX$_2$ in the presence of spin-orbit coupling. All energies are measured with respect to the Fermi energy. Note that the energy scales are different for each plot. }
\end{figure*}

We now discuss the effects of the spin-orbit interaction on the electronic structure of the Sb(Mo) doped systems. The band structures in the presence of SOC are shown in Fig. \ref{fig:bands_sb_soc}. As we can see, the main effect of the interaction is the introduction of small splittings in the impurity levels, which are highlighted in orange and ordered following the notation introduced in Fig. \ref{fig:bands_sb}. In particular, the levels $E_2$ and $E_3$ remain near-degerate and split into two pairs by about $0.05$ eV, which is the largest splitting observed. On the other hand, no splitting is observed in the valence band and a very small splitting is observed in the conduction band, in contrast with the features observed in the pristine materials. Moreover, note that the nature of the band gap is also modified by the presence of the impurity. In Sb-doped MoS$_2$, the VBM is shifted to the $\Gamma$ point of the supercell BZ. In MoSe$_2$, the valence band energies at $\Gamma$ and $K$ are nearly identical and in MoTe$_2$ the VBM lies at the $K$ point. In all cases, the CBM lies at a point at the $\Gamma-K$ line, but note that the conduction band is particularly flat throughout a portion of the BZ in MoS$_2$ and MoSe$_2$. 

\begin{figure*}[ht]
    \centering
    \includegraphics[width=0.8\textwidth]{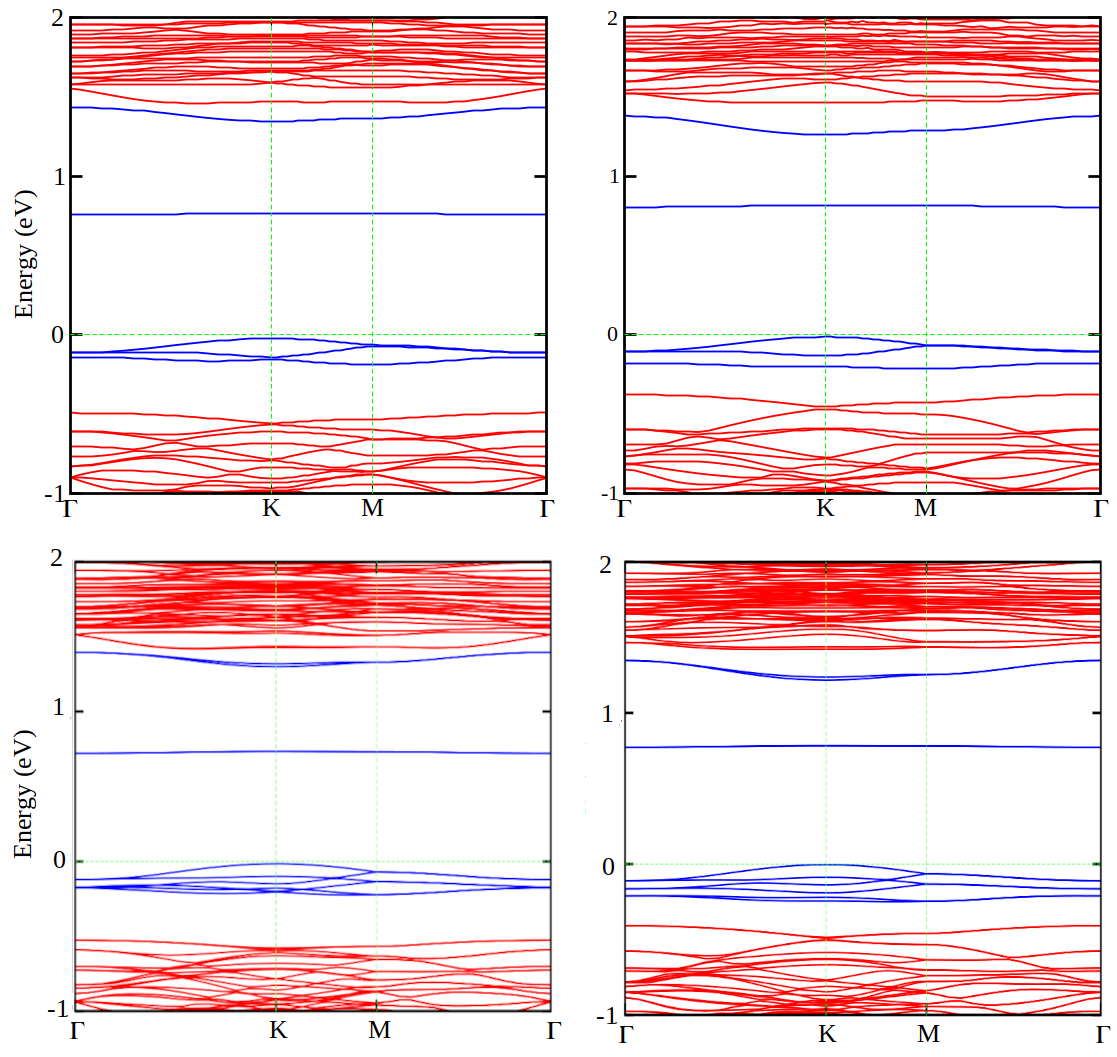}

    \caption{\label{fig:S11} Band structures of Sb(Mo) doped MoS$_2$ obtained from a calculation with a relaxation of atomic positions, maintaining the supercell vectors fixed (left panels) and a calculation with a full relaxation, including the supercell vectors (right panels). The top (bottom) panels show calculations without (with) spin-orbit coupling. Notice the effects of the full-relaxation on the band structure, particularly in the first set of impurity levels. These calculations were performed with the VASP code on a $5 \times 5$ supercell.}
\end{figure*}

The magnitude of the gaps are $1.77$, $1.58$ and $1.17$ eV for X = S, Se and Te, respectively. They are larger than the corresponding pristine gaps in the presence of SOC, in agreement with our calculations without SOC. This behavior contrasts with a previous calculation from Ref. \onlinecite{zhong2019electronic}, in which a slightly different band structure and a gap reduction were reported for X = S. We believe these differences are related to two factors.
First, a full cell relaxation may result in a slightly different band structure than one obtained in a calculation in which only internal degrees of freedom are relaxed. In Fig. \ref{fig:S11}, we display a comparison of these two situations both in the absence and the presence of SOC for Sb-doped MoS$_2$ (calculated with the VASP code). As we can see, the calculation without a full-relaxation, but with SOC agrees quite well with the band structure reported in Ref. \onlinecite{zhong2019electronic}.
Second, and more importantly, Ref. \onlinecite{zhong2019electronic} prescribes a different assignment for the valence band and the electronic structure of the impurity levels, despite the similarities to our results. In that work, the first bundle of six weakly-dispersive levels (shown in blue in Fig. \ref{fig:S11}) is assumed to consist entirely of extended states, such that the topmost band is assigned as the valence band. However, as we see in our analysis of the wavefunctions of these levels, shown in Figs. \ref{fig:bands_sb} and \ref{fig:ldos_sb} in the absence of SOC and Fig. S11 (supplemental material) in  the presence of SOC, these states, which correspond to $E_1$, $E_2$ and $E_3$, show a clear localization around the impurity so we assign them as impurity states. Additionally, as we have mentioned before, these results are consistent with those observed in a larger $10 \times 10$ supercell (see Fig. S5 of the supplemental material). Therefore, we consider that the valence band actually lies further below, corresponding to the closest red line in Fig. \ref{fig:S11}. 

In light of this discussion, if we use the valence band assignment of Ref. \onlinecite{zhong2019electronic}, we find band gaps between 1.4 and 1.5 eV for the calculations in Fig. \ref{fig:S11}. These are in good agreement with the theoretical value of 1.39 eV reported in that reference and would be consistent with a reduction when compared to the pristine values (see Table \ref{tab:structure_pure}). However, if we follow our assignment as described above, the actual gaps are around 1.8-1.9 eV for the same calculations and are consistent with an increase.
In that scenario, the redshifts of the A and B excitonic peaks observed in experimental photoluminescence measurements, which were previously assigned to a reduction of the band gap, could be related to other mechanisms. These include excitonic effects such as a different quasiparticle correction, a stronger electron-hole interaction in the presence of the impurity or a mixture of the original excitons with new transitions involving the impurity-induced levels. All these effects are not included in a ground state DFT calculation, so only a calculation that fully addresses them, such as a GW/BSE calculation, would be able to fully elucidate the optical properties of these materials.

Finally, we now briefly discuss other features that are not included in our calculations which could be relevant to the stability and the electronic properties of the substitutional impurities. First, note that we have only considered neutral charge configurations for all defect configurations. Previous studies of single vacancies with charge in MoS$_2$ \cite{komsa-prb-2015} indicate that, for S vacancies, the neutral state is favored for Fermi energy values (charge doping) throughout most of the gap, followed by a transition to a -1 charge state in the vicinity of the CBM, which is accompanied by the occupation of the associated defect levels. In contrast, for the Sb(S) substitutional defect, we see in our calculations that the defect levels are already (partially) occupied in the neutral state (Figs. S3 and S8 in the supplemental file), so a +1 charge state could be favored for Fermi energy values in the vicinity of the VBM. For Mo vacancies, Ref. \onlinecite{komsa-prb-2015} indicates the stability of different charge states throughout the gap, which is a consequence of the presence of multiple defect levels inside the gap. Considering that we see a similar defect level structure for the Sb(Mo) substitutional defect in our calculations, we expect a similar outcome for the stability of charged states. We stress however, that additional calculations would be required for a definitive conclusion, which are beyond the scope of the present work.

Second, since the electronic and optical properties of 2D materials can strongly be influenced by their environment, the substrate can play an important role in the description of the defect-induced levels inside the gap. For instance, Refs. \onlinecite{ugeda2014} and \onlinecite{raja2017} have found that the quasiparticle band gaps of semiconducting TMDCs can decrease by up to $\sim 100$ meV in the presence of capping graphene layers. As such, we expect that the defect levels may experience renormalizations of the same order.  On the other hand, the renormalizations due to quasiparticle and excitonic effects are found to partially cancel out, resulting in small changes to the optical properties. In Ref. \cite{zhong2019electronic}, the homogeneously Sb-doped MoS$_2$ nanosheet is prepared on top of SiO$_2$/Si substrate. Considering that SiO$_2$ is less polarizable than graphene, we expect that these renormalizations will be weaker in this case. However, the scenario may be very different with other 2D semiconductors, such as phosphorene, where much larger renormalizations are seen in the presence of a quartz substrate and capping layers of h-BN \cite{qiu2017}.

% ------------------------------------------------------------------------------- %

\section {\label{sec:conclusions} Conclusions}

In summary, we have studied the electronic and structural properties of substitutional Sb impurities in TMDCs of the form MoX$_2$, focusing on the Sb(Mo) substitution that was recently observed experimentally in MoS$_2$. First, we find that the Sb(Mo) substitution is more favorable than the Sb(S) substitution in MoS$_2$ under S rich conditions. This result supports a recent experimental observation in which this substitution is favored in homogeneously Sb-doped MoS$_2$ nanosheets grown by CVD, where temperatures as high as 955K are achieved and the gas phase of S$_2$ should be predominant over other S allotropes. Similar trends are found in defective MoSe$_2$ layers, but in MoTe$_2$ layers the Sb(Mo) is less likely to form under similar conditions considering that gaseous phases of Te are less likely to be present. Next, studying the electronic properties of the Sb(Mo) substitution, we find that, in the absence of spin-orbit coupling, five impurity levels are found inside the band gap, with energies that span the entire gap. The Fermi energy lies a few tenths of eV above the valence band edge, suggesting a predominant $p$-type behavior. Projected and local density of states calculations reveal that the wavefunctions associated with the levels have very distinct shapes. Only two non-degenerate levels include contributions from the $5s$ orbital of the Sb impurity, while the others only contain contributions from $4d$ orbitals of second-neighbor Mo atoms and small contributions from $p$-orbitals of first-neighbor X atoms. We verify that these properties are very similar to those found in defect levels induced by a single vacancy in MoS$_2$, suggesting that the defect levels found in the Sb(Mo) doped system result from a rehybridization of vacancy-induced levels and the $5s$ orbital of Sb. 

Finally, we find that the spin-orbit interaction induces splittings in the impurity levels that are roughly of the same order of those found in the valence and conduction bands of the pristine systems. Moreoever, according to our identification of the valence band and the impurity levels in the doped systems, the band gaps are larger than those found in the pristine materials. This suggests that other effects could be behind the experimentally observed redshifts of the excitonic A and B peaks, including excitonic effects such as a stronger electron-hole interaction in the presence of the impurity or a mixture of the original excitons with new transitions involving the impurity-induced levels. We stress however, that only a calculation that fully accounts for these effects, such as a GW/BSE calculation would be able to provide a complete picture.
Nevertheless, we believe that such a multi-level electronic structure enriches the optical response of these materials. The position of the impurity levels, and therefore the optical transition energies, could be tuned by the choice of material and, possibly, by the defect concentration. In that scenario, further investigations, particularly those focusing on the effects of disorder and the electron-hole interaction may continue to reveal the fundamental properties and potential applications of these materials in future optoelectronic devices.

\begin{acknowledgements}
We thank CNPq, CAPES, FAPERJ and INCT Carbon Nanomaterials for financial support. We also thank LNCC-MCTI and NACAD-UFRJ for the computational resources employed in this work. SU also thank DEAC at  Wake Forest University for providing additional computational resources. Finally, we also thank Rodrigo B. Capaz for valuable discussions.
\end{acknowledgements}

Marcos G. Menezes, and Saif Ullah contributed equally to this work.

\nocite{*}

\bibliography{bibliography}% Produces the bibliography via BibTeX.

\end{document}